# Automatic Generation of the Axial Lines of Urban Environments to Capture What We Perceive

Bin Jiang and Xintao Liu

Division of Geomatics, Department of Technology and Built Environment
University of Gävle, SE-801 76 Gävle, Sweden
Email: bin.jiang@hig.se, xintao.liu@hig.se

**Abstract**
Based on the concepts of isovists and medial axes, we developed a set of algorithms that can automatically generate axial lines for representing individual linearly stretched parts of open space of an urban environment. Open space is the space between buildings, where people can freely move around. The generation of the axial lines has been a key aspect of space syntax research, conventionally relying on hand-drawn axial lines of an urban environment, often called axial map, for urban morphological analysis. Although various attempts have been made towards an automatic solution, few of them can produce the axial map that consists of the least number of longest visibility lines, and none of them really works for different urban environments. Our algorithms provide a better solution than existing ones. Throughout this paper, we have also argued and demonstrated that the axial lines constitute a true skeleton, superior to medial axes, in capturing what we perceive about the urban environment.

## 1. Introduction

Open space of an urban environment or of a city is the space where people can freely move around, and it is in contrast to closed spaces such as buildings and street blocks. Open space can be thought of as something between closed spaces, and can be represented as a complex polygon with holes. The holes denote closed spaces, i.e., the buildings and street blocks. Open space of a city forms one continuous piece rather than separate parts, since people can travel from anywhere to anywhere else in a city. For the purpose of modeling human movement, open space has been of primary interest, being the modeling target in space syntax (Hillier and Hanson 1984). It is done through representing open space as many perceivable units, and further forming a connectivity graph consisting of nodes representing individual units and links if the units are intersected. The connectivity graph can be used to analyze and rank the status of the individual units for understanding the underlying topology of the city. The representation of open space can be in different ways.

One classic representation is to represent open space as many interconnected axial lines, each depicting a linearly stretched part of open space. The axial lines constitute what is often called an axial map that can be thought of as a skeleton of the city. The axial map is supposed to consist of the least number of longest axial lines. Its generation is achieved by manually drawing the first longest visibility line, followed by the second longest, the third longest, and so on, until the entire open space is covered with interconnected axial lines. The drawing procedure can be justified by the partitioning (or discretization) of a large scale space (i.e., the continuous open space) into numerous discrete small scale spaces, each of which is represented by an axial line – the longest visibility line for the small scale space. A major difference between a large scale space and a small scale space lies in whether or not it is perceivable from a single vantage point, no for a large scale space and yes for a small scale space. The notions of large scale and small scale spaces constitute the foundation of our proposed algorithms for automating the axial lines or the axial map.

Space syntax has been widely applied to various urban related studies such as traffic modeling for pedestrians and vehicles, crime analysis, and human wayfindings in complex urban environments (Hillier 1996). Such studies significantly rely on the axial maps, which are currently created by hands.

The hand-drawn map has been a very controversial aspect of space syntax (e.g., Jiang and Claramunt 2002, Ratti 2004). Because of this, many alternative spatial representations have been suggested, for instance, visibility graph (Turner et al. 2001), point-based space syntax (Jiang and Claramunt 2002), and more alternative representations based on street networks (e.g., Jiang and Claramunt 2004, Thomson 2003, Jiang, Zhao and Yin 2008). In the meantime, efforts have been made towards automatic generation of the axial map in the space syntax community (c.f., next section for an overview). However, despite these efforts, few solutions can produce the axial map that consists of the least number of longest visibility lines (or axial lines), and none of them really works for different urban environments. In this paper, we provide a solution that can automatically generate the axial map that consists of the least number of longest axial lines, thus making space syntax analysis more objective and more meaningful for urban morphological studies. The contributions of this paper are two-fold, which are reflected in the paper title. First, it provides a simple yet elegant solution to the automation of the axial lines through the novel concept of bucket. Second, it discovers an important statistical difference between the medial axes and the axial lines of urban environments, i.e., the axial lines can much better then medial axes capture what we perceive about urban environments.

This paper is structured as follows. The next section presents an overview about axial lines related concepts, and various efforts towards automation of the axial map. Section 3 describes in detail the major motivation, ideas and pseudo codes of our algorithms, which are implemented as a research prototype. Section 4 reports our primary experiments of different urban environments and assesses a major statistical difference between medial axes and axial lines. Finally section 5 concludes and points to further work in the future.

**2. Axial lines related concepts and efforts towards an automatic solution**
Both axial lines and medial axes (Blum 1967; Blum and Nagel 1978) are considered to constitute a skeleton of urban environments, but they differ fundamentally. Take a rectangle shaped space for example (Figure 1a), where medial axes are indicated by the dashed lines. The medial axes or skeleton can be thought of as the loci of centers of bi-tangent circles that fit entirely within a shape. The skeleton can also be produced by the thinning operation in image processing. The medial axes have a nature of sensitivity, i.e., a slight change in the shape can cause a dramatic change to its skeleton (Figure 1b and 1c). Due to this sensitivity, the skeleton based on the medial axes has difficulty in forming a natural decomposition of a shape into a set of basic units that mirror the shape we perceive (Katz and Pizer 2003). For example, despite of two tiny convex and concave parts, both shapes shown in Figure 1b and 1c are perceived as a rectangle. However, the perceived shape rectangle can hardly be seen from their medial axes. Medial axes can only be effective in capturing the parts we perceived for purely symmetric objects or shapes, and they become defective if they are slight asymmetric. This can also be seen from the medial axes of the Gassin open space (c.f. Figure 4 later in the paper), where each hole is surrounded by a set of precisely defined medial axes. In contrast, the "roughly" defined axial lines can capture well the parts we perceive - a few longest lines intersected with many shorter ones. It is in this sense that the axial map is a true skeleton of a city, reflecting what we perceive. This point will become clearer when we further illustrate a major statistical difference between medial axes and axial lines in the following experiments.

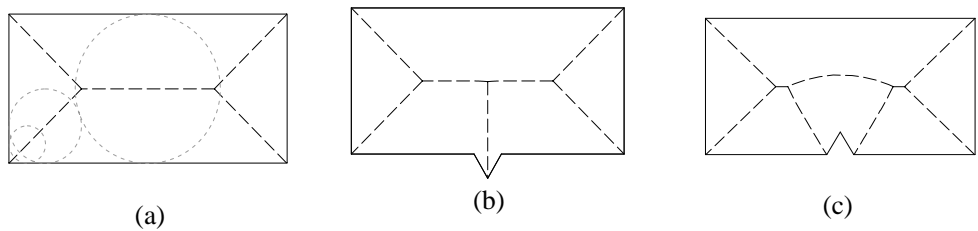

(a)      (b)      (c)

Figure 1: Illustration of Blum's medial axes and their sensitivity

The initial definition of axial map, given by Hillier and Hanson (1984), is based on a prior definition of the convex map of open space – a complex polygon with holes; Refer to Figure 2 for an illustration of various concepts used in the paper. The convex map is "the least set of fattest spaces that covers the system" (p. 92). The system here is what we refer to as open space. The procedure of generating the convex map seems based on some arbitrary rule, i.e., "simply find the largest convex space and draw it in, then the next largest, and so on until all the space is accounted for" (p. 98). As a reminder, the space here is what we refer to as open space. Given the definition of convex map, an axial map is defined as "the least set of such straight lines which passes through each convex space and makes all axial links" (p. 92). This definition is a bit vague, for instance, what does it mean by "passes through each convex space"? However, the procedure of generating the axial map is less vague, similar to the one for generating the convex map, i.e., "first finding the longest straight line that can be drawn …, then the second longest, and so on until all convex spaces are crossed and all axial lines that can be linked to other axial lines without repetition are so linked" (p 99). This procedure of generating axial map seems much clearer than in its definition. However, there is no unique partition of a 2D open space into convex polygons, as argued by many researchers, in particular those working in the famous art gallery problem (O'Rourke 1987). Therefore, there are difficulties to draw the least number of axial lines that cover the convex polygons, according to the existing definition. Another related linear representation of open space is called all-line map. It consists of all the lines generated from the process where every pair of the vertices of the holes are linked and extended without intersecting the holes of open space. Without reference to the convex map, the all-line map can be objectively generated, and it has been used in space syntax analysis (Hillier 1996). Attempts have been made towards an automatic derivation of the axial map from the all-line map with a considerable degree of redundancy of lines.

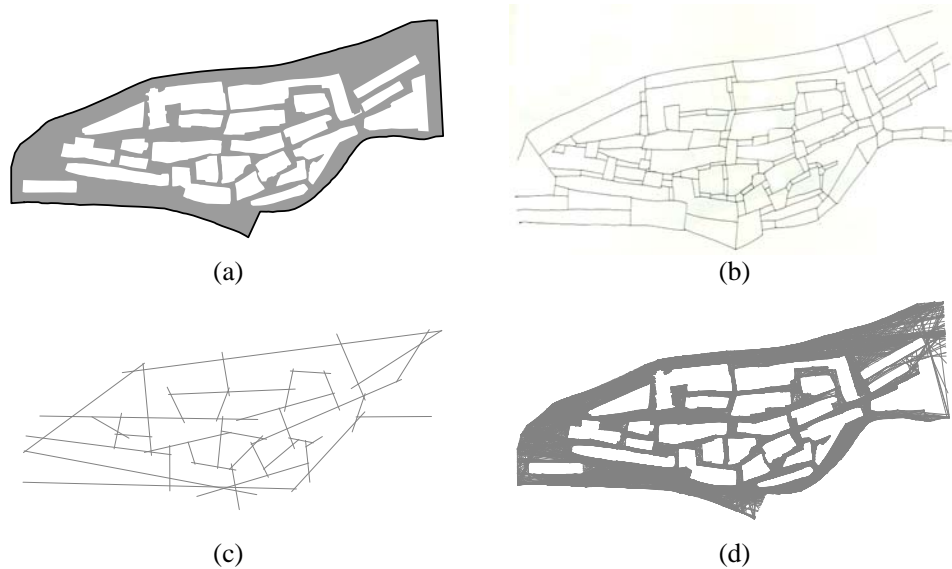

(a)　　　　　　　　　　　　　　　　　　(b)

(c)　　　　　　　　　　　　　　　　　　(d)

Figure 2: Illustration of various concepts based on the town of Gassin, (a) open space – the shaded part, closed spaces – the white parts, and outmost boundary – the black line, (b) original convex map, (c) original axial map and (d) all-line map produced from Depthmap,

Peponis et al. (1997, 1998) suggested an alternative space partition called s-partition, which can be automatically and uniquely generated. The spaces formed by s-partition are called s-partition spaces. Peponis and his colleagues took the s-partition spaces as a sort of convex spaces, and used them as a reference to derive an economic set of lines from the all-line map. Their procedure starts with picking up the first axial line with the most crossing points with the underlying s-partition spaces. It continues to the second axial line which crosses the second most number of s-partition spaces (not crossed by the first axial line), and so on, until no more s-partition space left. In essence, it is a greedy strategy, with a local optimum to achieve a global optimum. It sounds like an elegant algorithm, but the procedure does not always lead to the axial map, even with some remedy to the procedure. This is the very reason that Peponis et al. avoided the use of axial lines and called them m-lines. Besides, this procedure has a problem of resolution, i.e., open space must be significantly simplified before the method can be applied.

Following up the pioneering work by Peponis et al, Turner et al. (2005) developed a procedure to remove the constraint of resolution and suggested another subset elimination technique for reducing the all-line map into the axial map. In the course of the subset elimination, Turner et al.'s algorithm considered two constraints: surveillance and topological rings. The surveillance refers to the fact that an entire open space is under surveillance of all the axial lines. If any part of open space is not under the surveillance, then new axial lines are needed. Every hole is supposed to be surrounded by a topological ring. In case the topological ring is incomplete, new axial lines are again needed to complete it. Despite the improvement, there are still some pathological cases that cannot be solved by the existing solution, as indicated by Turner and his colleagues. More importantly, the algorithm does not always produce the least number of longest axial lines for any urban environment, implying some redundant lines existed.

The previous efforts led by Peponis and Turner stick too dogmatically to the initial definition based on convex partition, yet reflect Hillier and Hanson's rules rather than what we perceive. Nevertheless, they provided the best solutions so far in terms of producing the axial map that consists of the least number of longest axial lines. Other efforts do not lead to or target to the axial map. For instance, inspired by the notion of isovists (Benedikt 1979), which is the part of open space that is directly visible from a location, Batty and his colleagues (Batty and Rana 2004; Carvalho and Batty 2005) believed that the dominant direction of an isovist (so called isovist ridge) can capture a linearly stretched part of open space, and experimented on alternative representations. Unfortunately, the alternative representations they explored are far from the axial map; they avoided this problem and suggested instead to reformulate space syntax because of the impossibility of automatically generating the axial map. Contrary to these efforts, we will illustrate that an automatic solution to the axial map is not only possible, but also algorithmically sound.

**3. A solution for automating the axial lines based on the concepts of isovists and medial axes**
From the previous section, we can see that skeletonization (either the medial axes or the axial lines) is a process that discretizes a continuous open space into many individual lines. The medial axes are composed of numerous medial points of corresponding edges of holes, including also the outmost boundary. Obviously, the medial axes lines could be curved if the corresponding edges do not parallel each other. When it comes to axial lines, the discretization process is not as simple as that of the medial axes. We propose a two-stage solution to this problem (Figure 3). We first discretize open space into numerous rays that significantly overlap each other (i.e., the rays fully filled in open space), and then reduce the rays into an economic set representing the axial lines. Our innovation lies in a simple yet elegant reduction algorithm based on a novel concept of bucket that we developed.

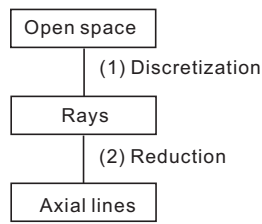

Figure 3: A two-stage solution to generating axial lines

The numerous overlapped rays fully fill the continuous open space. Putting it differently, all parts of open space are crossed by the rays, and all the rays that can be linked to other rays **with** repetition are so linked. This statement sounds like that in the procedure of generating axial lines, "first finding the longest straight line that can be drawn ..., then the second longest, and so on until *all convex spaces are crossed and all axial lines that can be linked to other axial lines without repetition are so linked*" (p 99, emphasis by the authors). The only difference lies in the fact that the rays are repeated in the sense that some rays can be replaced by others, while the axial lines are without repetition, implying that every axial line is unique. In other words, the rays constitute a super set of the axial lines. Through elimination of repeated ones we can obtain an economic set of the axial lines - the least number of longest visibility lines of open space. Now let us describe in details how the two stages can be achieved.

It is a pretty straightforward process to generate a full set of rays that entirely cover open space. For each vertex of the medial axes, there is a ray, i.e., the dominant direction of the vertex's corresponding isovist, or simply called an isovist ridge (Carvalho and Batty 2005). The process of generating rays must be exhausted for all the vertices of the medial axes. We also tried to generate a full set of rays by linking every pair of the vertices of the medial axes, but eventually we abandoned this idea because of the intensiveness of the computation. The generated rays are significantly redundant or repeated. To eliminate the repeated rays, we introduce a new concept of bucket. Bucket is defined as a space that is approximated by an axial line, and it is represented as a polygon that surrounds an axial line or a ray (Algorithm I). To remind, every vertex of a medial axis has two associated points (in its corresponding shape edges) from which the vertex is created. This can be clearly seen from Figure 1, where the central points of the circles are the medial axes vertices created from two associated points in the rectangle edges that are tangent to the circles. For a complex polygon with holes, all medial axes vertices are created from the edges of holes or the outmost boundary. Thus every medial axes vertex has two associated points in the edges of holes or in the outmost boundary. This can be equally said that every point in the edges of holes or the outmost boundary has one associated medial axes vertex.

It is through such an association that we derive a bucket for each ray. For example, the ray shown in red in Figure 4c cuts across three Voronoi regions (which are formed by medial axes) of three closed spaces A, B and C. More specifically, the ray with two ending points $e_1$ and $e_2$ cuts the Voronoi regions at vertices $x_1$ and $x_2$. According to the association mentioned above, we can find 4 associated points for the ending points, i.e., $e_{11}$ and $e_{12}$ for $e_1$, $e_{21}$ and $e_{22}$ for $e_2$. In addition, the ray segment $x_1x_2$ corresponds to two medial axes segments $x_1y_1$ and $x_2y_1$ that intersect at point $y_1$. The point $y_1$ has two associated points $y_{11}$ and $y_{12}$. The two ending points and the associated points can be chained together to form the bucket of the ray. In more general, the points to be chained together are $e_1, e_2, e_{11}, e_{12}, e_{21}, e_{22}, y_{11}, y_{12}, \ldots y_{n1}, y_{n2}$, where n is the intersection points of the corresponding medial axes segments. According to our observation, there are three intersection points in maximum for the medial axes segments. The number of the intersection points could be 0, implying that there is just one medial axes segment corresponding to the ray segment.

The bucket will help us to remove those redundant or repeated rays. The process can be described as follows. The first axial line is obtained by sorting all the rays, and selecting the longest. All other rays within the bucket of the axial line will be removed from the ray set, for they are not long enough, violating what we said the longest visibility lines. This process goes on for the remaining ray set for the second and third longest axial lines and so on, until there is no ray left in the ray set. In the end, we generate a set of axial lines, containing all the longest lines. This is a loop procedure using a global search strategy. The procedure can also be conducted in a recursive fashion and using a local search strategy. The procedure starts with the first longest axial line from the ray set. The second axial line is picked up from those rays directly connected to the first axial line, rather than from all the remaining rays, making it a local search strategy. The second axial line is not necessary to be the second longest at the global scale. This is the major difference between local search and global search: see more details in Algorithm II. Up to this point we have generated an economic set of the axial lines that are connected and cover all parts of open space. It is important to note that the reduction procedure does not change the nature of all the rays, except the removal of redundant or repeated rays.

With Algorithm II, we generate all the rays from the medial axes vertices for open space and then reduce them into an economic set representing the axial lines. Algorithmically, we could do this in a recursive fashion from the very beginning. We start with an arbitrarily drawn line in open space, and discretize the line into a series of points. From the points, we generate isovist ridges, and remove those repeated ones using the bucket reduction algorithm. For those selected lines, we recursively repeat the above process; see details in Algorithm III. After the recursive process for all locally generated rays, we may need another round of reduction to clean a few lines that are still considered to be redundant, to achieve the least number. This algorithm sounds very simple, but there is a resolution issue involved in the process. We adopt a resolution that is determined by one third of the smallest distance between two closed holes or between any hole and the outmost boundary. The resolution is proved to be sufficient for generating the axial map.

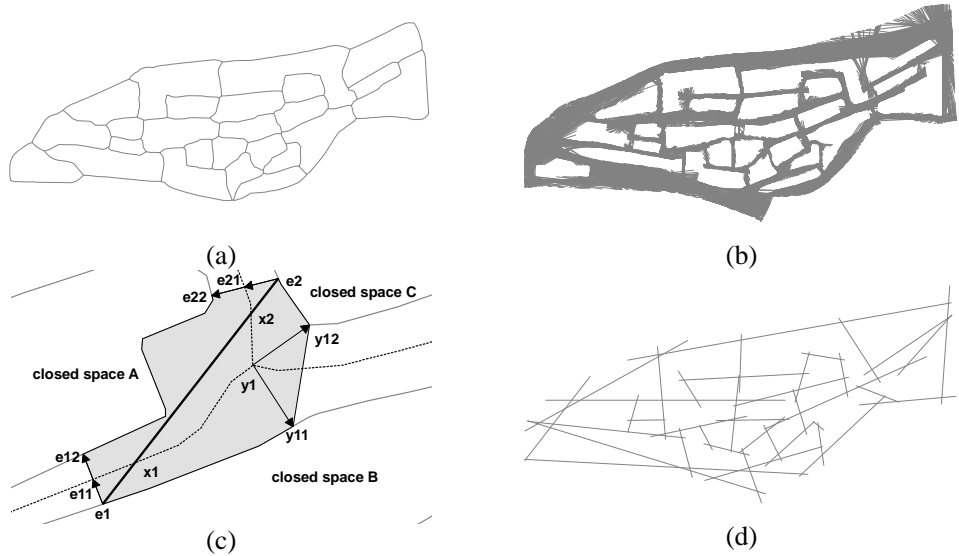

Figure 4: Illustration of the solution from (a) medial axes, (b) all the rays generated individual vertices of the medial axes, (c) bucket of a ray, and (d) generated axial map
(Note: for panel c: thick line = ray, dashed lines = medial axes, gray lines = edges of closed spaces, arrows = association, the boundary of shaded area = bucket generated for the ray)

**Algorithm I: Procedure for extracting a bucket of a ray or an axial line**
```
------------------------------------------------------------------------
Input: A ray r and a complex polygon with holes
Output: The bucket of the ray r

Function BucketFormation (ray r)
    Generate Voronoi regions or medial axes of the complex polygon;
    Intersect current ray with the Voronoi regions to obtain partitioned ray segments with
        intersection points x1, x2, … xm;
    For the ending points e1 and e2, find the associated points: e11, e12 for e1, and e21,
        e22 for e2;
    For every partitioned ray segment, find the associated medial axes segments, their
        intersection points y1, y2,. . . yn (n<=3), and the associated points: y11, y12 for y1,
        y21, y22 for y2, and y31, y32 for y3;
    The two ending points and the associated points are chained together to form a closed
        polygon called bucket of the ray.
```

**Algorithm II: Loop and recursive procedures of generating axial lines globally**
```
------------------------------------------------------------------------
Input: Open space S
Variables: Ray set R
Output: Axial line set A

// Generate ray set R for the entire open space from the vertices of medial axes (MA)
```

```
Function Globally_Generate_Ray_Set ()
    Generate medial axes for each closed space;
    For each medial axis
        For each vertex of the medial axis
            Generate isovist ray set I at every one degree direction (0 ~ 360⁰);
            Select the longest ray (l_ray)with the best connectivity as ridge from isovist
                ray set I;
            Add ray to ray set R;

// Loop version of reduction process
Loop Function Global_Search ()
    Select the longest ray from the ray set R as current axial line in the axial line set A;
    While (the ray set R is NOT empty?)
        Construct the bucket of current axial line;
        If (any ray in the bucket?) Then
            Delete this ray from the ray set R;
        Select the longest ray from the remaining ray set R
            as another axial line in axial line set A;

// Recursive version of reduction process
Select the longest ray from the ray set R as current axial line in the axial line set A;
Recursive Function Local_Search (current axial line)
    BucketFormation(current axial line);
    If (any ray in the bucket?) Then
        Delete this ray from the ray set R;
    Select a subset of rays intersecting the current axial line;
    If (the subset is NOT empty?) Then
        Select the longest ray in the subset as an axial line in
            the line set A;
        Local_search (an axial line);
```

**Algorithm III: Recursive procedure of generating axial lines locally**
-----------------------------------------------------------------------------------------
```
Input: Open space S
Variables: Ray set R
Output: Axial line set A

// Generate ray set R from individual points of a ray
Arbitrarily draw a line within open space S, and extend it as a current ray;
Recursive Function Locally_Generate_Ray_Set (current ray)
    Generate isovist ridges list along the current ray;
    Current ray = the longest ridge
    BucketFormation (current ray);
    For each ray in isovist ridges list
        If (the ray in accumulative bucket) Then
            Delete this ray from ridges list;
    If (the remaining ridges list is NOT empty?) Then
        Add remaining ray in ridges list to ray set R;
        For each ray in remaining ridges list
            Locally_Generate_Ray_Set (ray);
```

The above algorithms can be justified by the fact that each axial line is a small scale space that is represented by a bucket, and all those rays within the bucket are considered to be redundant and shall therefore be removed. The key of the algorithms lies in the partitioning of a large scale open space into many small scale spaces, each of which is represented by a bucket. The bucket is not necessarily convex, so it is not a small scale space per se. However, the bucket can roughly be considered to be a small scale space, since it is visible entirely from an axial line point of view. The set of buckets constitutes the least number of spaces that cover all parts of open space. Above algorithms have been put together as a research prototype namely AxialGen (Jiang and Liu 2009), which has been released for academic uses.

## 4. Experiments with a set of urban environments

We made a series of experiments with a set of test cases in order (1) to demonstrate how able the axial lines are in capturing units or parts we perceive about urban environments, and (2) to assess the validity of the axial lines and computational complexity. Some of the test cases are taken from the space syntax literature, while others are some typical street patterns (Jacobs 1995). As we can see from Figure 5, all axial maps generated, including a few pathological cases identified previously in the literature, consist of the least number of the longest visibility lines. The computation of an axial map involves the derivation of the medial axes, the rays that cover the entire open space, and the reduction of the rays into an economic set being the axial lines. With the current implementation, the medial axes are created using the operation of distance transformation or distance map, which can be efficiently done with most geographic information systems software. We adopt binary space partitioning that can greatly speed up the process of generating the rays (Sung and Shirley 1994). In terms of computational complexity, the most time consuming part is attributed to the reduction process, where ray-in-polygon operation consumes too much computational resource. The computation is still deemed very intensive (Table 1), but the axial lines generated are nearly perfect. All the axial lines shown in both figure 5 and figure 6 are automatically generated using the local search option of Algorithm II. It should be noted that with the global search option of Algorithm II or Algorithm III, it leads to a similar perfect result, given that the parameters are appropriately set.

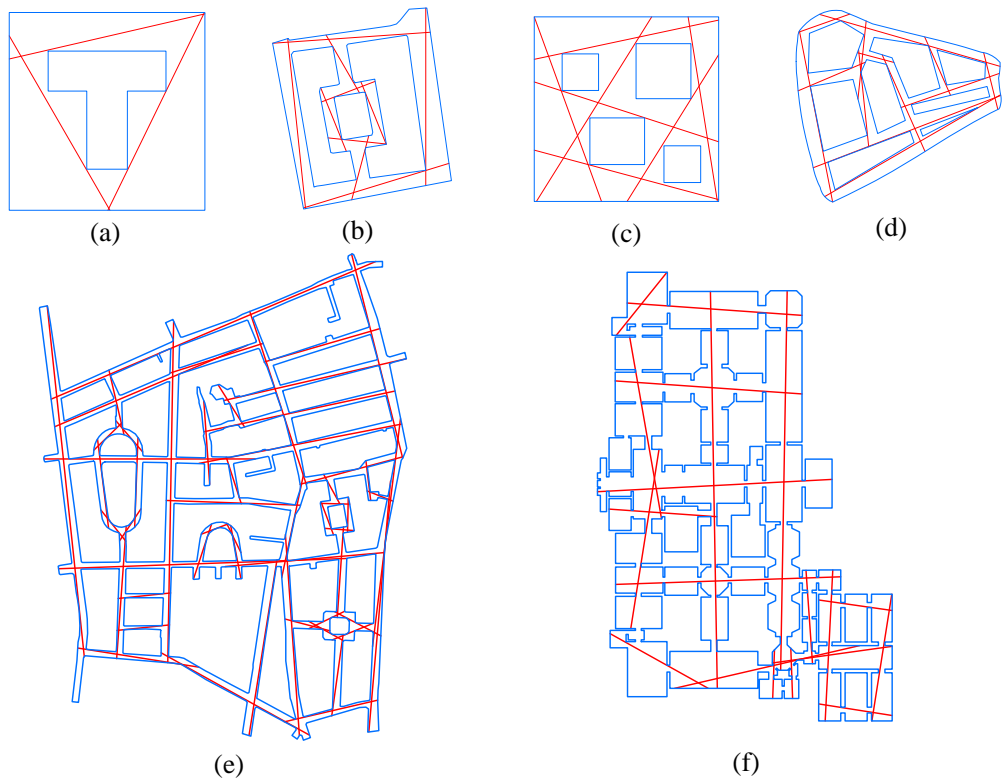

Figure 5: (Color online) The first group of experiments of urban environments taken from the space syntax literature: (a) T-shape, (b) Three-block case, (c) Four-block case, (d) Eight-block case, (e) Barnsbury, and (f) National Gallery London (rotated by 90 degrees)

We can also note from the experiments that the axial lines, rather than the medial axes, are able to capture the individual parts we perceive about an urban environment towards an understanding of underlying urban structure, i.e., a very few long axial lines interconnected to many shorter ones. Usually, the number of the axial lines is far less than that of the medial axes (Table 1). Importantly the axial lines demonstrate a hierarchy, which is a key feature of complex systems (Simon 1962), and that of urban systems as well (Salingaros 2005, Jiang 2008). This hierarchical structure is clearly absent in the skeleton representation by the medial axes. Taking the case of Paris as an example, its medial axes all have a similar length, exhibiting a Gaussian like distribution, whereas its axial lines bear a power law like or heavy tail distribution (Figure 7). The axial map captures the parts we can conceive or what we truly perceive, being the image of the city (Lynch 1965). With the axial map, a few longest lines, which are visually dominating, constitute the landmarks that can be distinguished from all other axial lines. This image of the city can further be seen from the axial maps visualized (Figure 8) according to local integration, one of the key space syntax metrics (e.g., Jiang 2006, Jiang 2007), with red representing the most integrated lines, and blue the most segregated ones. The reader can assess visually how the kind of hierarchical structure appears within the different urban environments.

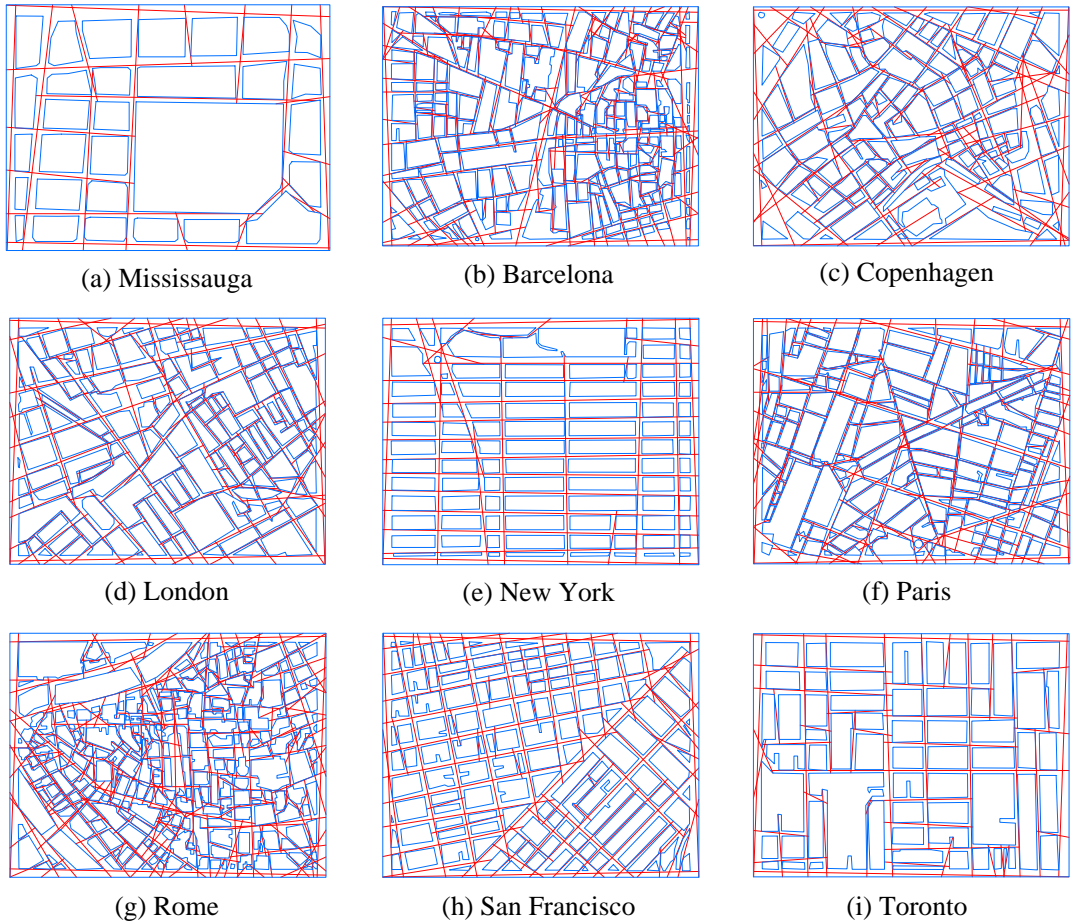

Figure 6: (Color online) The second group of experiments using the typical street patterns

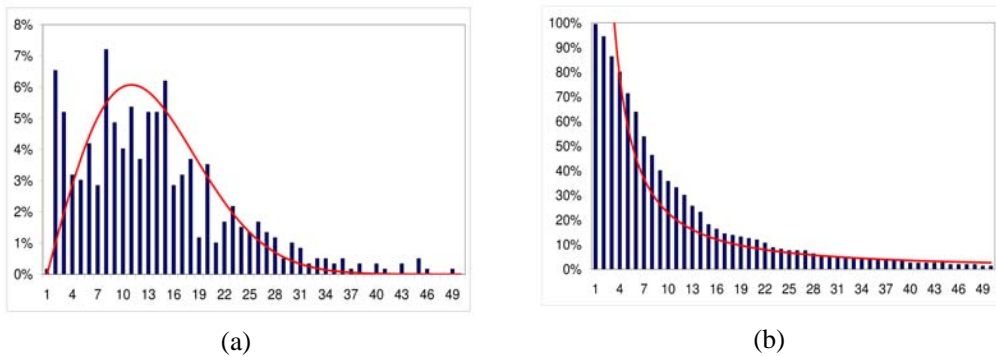

Figure 7: (Color online) A major statistical difference between the medial axes and the axial lines: (a) Non-hierarchical structure with the medial axes whose length exhibits a Gaussian like distribution, and (b) Hierarchical structure with the axial lines whose length demonstrate a power law like or heavy tail distribution.

Table 1: Time cost for generating the axial lines and the number of the axial lines and that of the medial axes
(The computation was done on a desktop computer: RAM: 4 GB, CPU: 2.4GB, 4 cores, Hard disk: 250GB, OS: Windows XP)

| Patterns | Time cost (seconds) | Medial axes | Axial lines |
|---|---|---|---|
| Barnsbury | 320 | 103 | 55 |
| Gassin | 90 | 72 | 39 |
| Mississauga | 77 | 81 | 23 |
| Barcelona | 1611 | 689 | 217 |
| Copenhagen | 509 | 381 | 115 |
| London | 967 | 437 | 134 |
| New York | 2476 | 262 | 22 |
| Paris | 1594 | 597 | 160 |
| Rome | 1440 | 797 | 254 |
| San Francisco | 658 | 396 | 67 |
| Toronto | 227 | 175 | 46 |

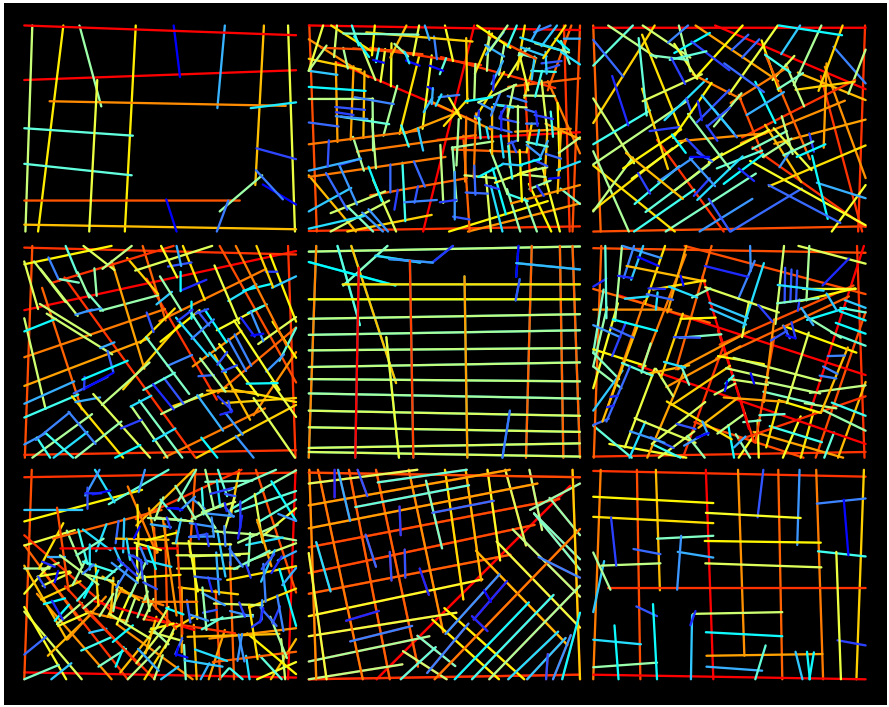

Figure 8: (Color online) Axial maps visualized according to local integration for showing underlying hierarchical structure of the street patterns

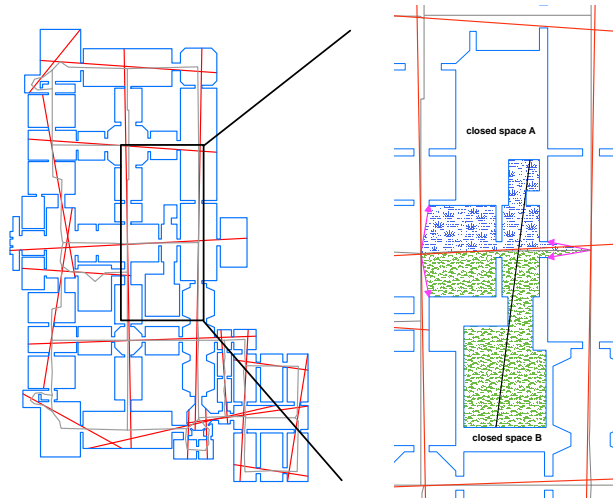

Figure 9: (Color online) A missing axial line (black) with National Gallery London (rotated by 90 degrees)

Despite the advantages we have shown, our algorithms are not without its problems. The bucket algorithm works only for linearly stretched parts that are formed by two closed spaces. The reader may have noticed already that there are a few axial lines missing from the experimented cases. The missing lines occurred in places where one closed space has a concave part, forming a linearly stretched space penetrating deep into the closed space. Figure 9 shows such a missing axial line in the case of National Gallery London. This problem happens in particular when two concave parts, respectively from two closed spaces, form a linearly stretched space penetrating in two opposite directions deep into the two closed spaces. Strictly speaking, the linearly stretched part is indeed between two closed spaces, but it is actually created by each closed space. This is indeed a constraint of the current bucket algorithm. A tentative solution is to detect those concave parts, and add possible missing lines. However, this remedy procedure may have some side effects, for instance, slowing down computational process.

## 5. Conclusion

The least set of longest visibility lines or the axial lines that cover entire open space of an urban environment constitutes the backbone of the environment and forms what we perceive about open space. We have argued and demonstrated throughout the paper that the axial lines are superior to the medial axes in skeleton extraction. We developed a set of algorithms that can automatically generate the axial lines. We abandoned the initial idea of deriving the axial lines based on the convex partitions, and adopted instead the idea that an axial line is simply a small scale space that can be represented by a bucket. The concept of bucket is probably the most novel contribution of this paper.

As tested with various urban environments, the generated axial lines can uniquely represent their individual small scale spaces. Our algorithms can deal with all the pathological cases identified and discussed previously, thus being more robust than existing ones. However, further research is still needed in the future. How to speed up the automating process remains a critical issue, in particular when generating the axial lines for an urban system involving thousands of buildings or street blocks.

**Acknowledgement**
Part of the research was carried out while the authors worked with the Hong Kong Polytechnic University. Thanks to Alasdair Turner for kindly sharing the datasets of Gassin, Barnsbury and National Gallery London, and to the referees for their constructive comments. We also thank Junjun

Yin for his assistance in digitizing the maps of the street patterns, and Petra Norlund for polishing up our English.**References:**
Batty M. and Rana S. (2004), The automatic definition and generation of axial lines and axial maps, *Environment and Planning B: Planning and Design*, 31, 615 – 640.
Benedikt M. L. (1979), To take hold of space: isovists and isovist fields, *Environment and Planning B: Planning and Design*, 6, 47 – 65.
Blum H. (1967), A transformation for extracting new descriptors of form, In: Whaten-Dunn W. (ed.), *Models for the Perception of Speech and Visual Form*, MIT Press: Cambridge, MA, 362 – 380.
Blum H. and Nagel R. N. (1978), Shape description using weighted symmetric axis features, *Pattern Recognition*, 10, 167 – 180.
Carvalho R. and Batty M. (2005), Encoding geometric information in road networks extracted from binary images, *Environment and Planning B: Planning and Design*, 32, 179 – 190.
Hillier B. and Hanson J. (1984), *The Social Logic of Space*, Cambridge University Press: Cambridge.
Hillier, B. (1996), *Space Is the Machine: a configurational theory of architecture*, Cambridge University Press: Cambridge.
Jacobs A. B. (1995), *Great Streets*, MIT Press: Cambridge, MA.
Jiang B. and Claramunt C. (2002), Integration of space syntax into GIS: new perspectives for urban morphology, *Transactions in GIS*, 6(3), 295-309.
Jiang B. and Claramunt C. (2004), Topological analysis of urban street networks, *Environment and Planning B: Planning and Design*, 31, 151- 162.
Jiang B. (2006), Ranking spaces for predicting human movement in an urban environment, *International Journal of Geographical Information Science*, x, xx-xx, Preprint, arxiv.org/abs/physics/0612011.
Jiang B. (2007), A topological pattern of urban street networks: universality and peculiarity, *Physica A: Statistical Mechanics and its Applications*, 384, 647 – 655.
Jiang B. (2008), Street hierarchies: a minority of streets account for a majority of traffic flow, *International Journal of Geographical Information Science*, x, xx-xx, Preprint, arxiv.org/abs/0802.1284.
Jiang B., Zhao S. and Yin J. (2008), Self-organized natural streets for predicting traffic flow: a sensitivity study, *Journal of Statistical Mechanics: Theory and Experiment*, P07008, arXiv ePrint: 0804.1630.
Jiang B. and Liu X. (2009), AxialGen: a research prototype for automatically generating the axial map, Preprint, arxiv.org/abs/0902.0465.
Katz R. A. and Pizer S. M. (2003), Untangling the Blum medial axis transform, *International Journal of Computer Vision*, 55, 139 – 153.
Lynch K. (1960), *The Image of the City*, The MIT Press: Cambridge, Massachusetts.
O'Rourke J. (1987), *Art Gallery Theorems and Algorithms*, Oxford University Press: New York.
Peponis J., Wineman J., Rashid M., Kim S. H., and Bafna S. (1997), On the description of shape and spatial configuration inside buildings: convex partitions and their local properties, *Environment and Planning B: Planning and Design,* 24, 761 – 781.
Peponis J., Wineman J., Bafna S., Rashid M., and Kim S. H. (1998), On the generation of linear representations of spatial configuration, *Environment and Planning B: Planning and Design*, 25, 559 – 576.
Ratti C. (2004), Space syntax: some inconsistencies, *Environment and Planning B: Planning and Design*, 31, 487 – 499.
Salingaros N. A. (2005), *Principles of Urban Structure*, Techne: Delft.
Simon H. A. (1962), The architecture of complexity, *Proceedings of the American Philosophical Society*, 106(6), 467-482.
Sung K. and Shirley P. (1994), Ray tracing with the BSP tree, in: Kirk D. (ed. 1994), *Graphics Gems III*, Morgan Kaufmann: San Francisco.12